\documentclass[12pt,a4paper]{article}

\usepackage[utf8]{inputenc}
\usepackage[T1]{fontenc}
\usepackage{amsmath,amssymb,amsfonts}
\usepackage{graphicx}
\usepackage{booktabs}
\usepackage{natbib}
\usepackage{hyperref}
\usepackage{color}
\usepackage{float}
\usepackage{caption}
\usepackage{subcaption}
\usepackage{array}
\usepackage{multirow}
\usepackage{rotating}
\usepackage[margin=2.5cm]{geometry}

\setcitestyle{numbers}

\title{Economic impact of biomarker-based aging interventions on healthcare costs and individual value}
\author{Federico Felizzi}
\date{\today}

\begin{document}

\maketitle

\begin{abstract}
We investigate the economic impact of controlling the pace of aging through biomarker monitoring and targeted interventions. Using the DunedinPACE epigenetic clock as a measure of biological aging rate, we model how different intervention scenarios affect frailty trajectories and their subsequent influence on healthcare costs, lifespan, and health quality. Our model demonstrates that controlling DunedinPACE from age 50 onwards can reduce frailty prevalence, resulting in cumulative healthcare savings of up to CHF 131,608 per person over 40 years in our most optimistic scenario. From an individual perspective, the willingness to pay for such interventions reaches CHF 6.7 million when accounting for both extended lifespan and improved health quality. These findings suggest substantial economic value in technologies that can monitor and modify biological aging rates, providing evidence for both healthcare systems and consumer-focused business models in longevity medicine.
\end{abstract}

\section{Graphical Abstract}

\begin{figure}[ht]
\centering
\includegraphics[width=0.85\textwidth]{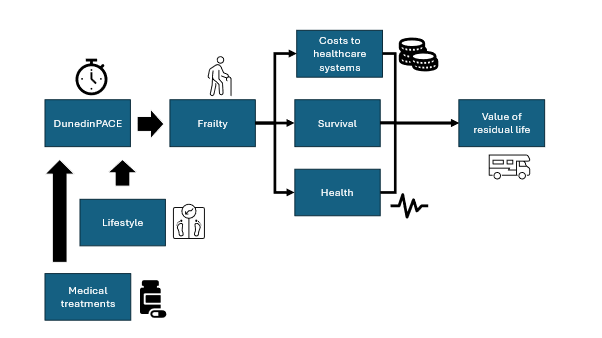}
\caption{Illustration of the dynamic influence of DunedinPACE, as controlled by novel health technologies such as pharmaceutical interventions and lifestyle changes. The modifications in DunedinPACE directly affect levels of frailty within the population. As frailty decreases, there are consequent improvements in three critical areas: direct healthcare system costs, survival rates, and overall health quality. These improvements collectively enhance the individual value of residual life, reflecting the economic and personal benefits of effective health management strategies. This model underscores the potential of targeted interventions in DunedinPACE to significantly alter health trajectories and economic outcomes.}
\label{fig:graphical_abstract}
\end{figure}
\newpage
\section{Introduction}
Population aging presents unprecedented socioeconomic challenges worldwide\cite{Ruckstuhl2023}, while simultaneously creating new opportunities for artificial intelligence-driven digital health solutions focused on the aging process. This study investigates potential business models that could address challenges related to aging and longevity medicine in real-world settings. We explore whether a combination of treatments, monitoring programs, and lifestyle interventions can meaningfully influence clinical outcomes related to aging and reduce costs associated with caregiving and diminished work capacity during reduced healthspan periods.

When establishing endpoints for longevity interventions, we draw upon Cummings et al.'s comprehensive framework\cite{Cummings2022}, which outlines various endpoints including total mortality, cause-specific mortality, and composite measures. We acknowledge the inherent challenges in establishing surrogacy between intermediate endpoints and final outcomes of interest, particularly regarding statistical power and necessary follow-up duration.

Among potential endpoints, frailty-related measures such as the Frailty Index\cite{Rockwood2005} show particular promise for our objectives because: (1) frailty demonstrates direct correlation with healthcare expenditures\cite{Hoogendijk2019}, and (2) numerous studies have established robust relationships between frailty, all-cause mortality, and major comorbidities including cancer, dementia, and cardiovascular disease.

A promising biomarker for measuring biological age is the DunedinPACE clock, which quantifies the pace of epigenetic change through DNA methylation patterns. Recent research demonstrates that DunedinPACE can predict frailty development and chronic disease onset more effectively than chronological age or alternative biomarkers\cite{Mak2023}. Importantly, this clock can detect impacts of early-life challenges, lifestyle modifications, and genetic factors on aging velocity, offering a comprehensive measure of biological age that reflects past exposures, present condition, and future health trajectories.

In this manuscript, we propose a novel biomarker-based monitoring and intervention strategy for longevity medicine based on routine DunedinPACE assessments in clinical settings. We hypothesize that measuring individuals' pace of aging enables identification of those at risk for accelerated aging, allowing for targeted preventive measures and personalized interventions to slow the aging process and improve health outcomes. We estimate the economic implications of this approach through a predictive model simulating different personalized recommendation scenarios designed to optimize epigenetic clock trajectories.

Our analysis establishes connections between DunedinPACE, frailty progression, and mortality outcomes to estimate:

\begin{enumerate}
    \item The impact of interventions (at varying efficacy levels) on total healthcare system costs
    \item Individual willingness-to-pay at different ages based on anticipated improvements in healthspan and lifespan
    \item Quality-Adjusted Life Year (QALY) improvements resulting from changes in quality-of-life utility and lifespan extension
\end{enumerate}

Through this approach, we aim to develop an economically viable business model that leverages digital technology and epigenetic monitoring to extend healthy lifespan through prevention and treatment of age-related conditions.

\section{Methods}

\subsection{Review of the endpoints used in longevity trials}

We began with a PubMed and Google Scholar review of the main trials used in Longevity Medicine, to find out the key endpoints that are relevant for our analytic model and decision-making framework. We did not do a systematic review, because the PICOS (Population, Intervention, Comparator, OutcomeS) criteria are very wide, so the number of hits might be too big, and therefore difficult to handle even for new AI-based solutions. Our target population includes all adults, with no age limit. The Intervention of interest is any drug, monitoring -- including digital health, and lifestyle change. The comparator is anything, which can mean either maintaining the current lifestyle or making it worse. The outcomes are what we are looking for. These can include endpoints such as death, the occurrence of specific diseases or quality of life questionnaires. The large scope of the search terms made us do a focused review, based on key studies that showed the endpoints that were used before in trials targeting longevity. So, to help us choose the trials and endpoints to consider, we used two reviews. One paper explains the pros and cons of the longevity-relevant endpoints, with the goal to provide complete recommendations on the trial designs in longevity and their relevant endpoints. It also talks about the potential surrogate endpoints, which can help us overcome the limitations of endpoints that need a very long time to produce a number of events that can eventually lead to the sufficient significance and statistical power to determine if the intervention included in the trial works\cite{Cummings2022}. Another more recent study builds on the previous work and proposes new biomarker-related endpoints that can serve as surrogate. The authors present a complete list of ongoing or finished trials, with interventions ranging from pharmacological to lifestyle changes and novel endpoints such as those linked to DNA methylation\cite{Moqri2023}.

\subsection{Frailty Index and its relationship to the prevalence of frailty}

The frailty index is a comprehensive metric designed to assess the vulnerability of older adults by quantifying their health status across a spectrum of clinical deficits. Typically, this index encompasses a wide range of health deficits---often 30 to 40---encompassing symptoms, diseases, disabilities, and laboratory abnormalities. Each deficit is evaluated, and the index is calculated by dividing the number of deficits present in an individual by the total number of deficits assessed, yielding a score between 0 and 1. A higher score indicates a higher degree of frailty.

The definition of frailty in a clinical context is based on established cutoff scores that determine a person's health classification. Typically, individuals with a frailty index of 0.25 or higher are labeled as "frail," signifying a high chance of negative health events. On the other hand, an index ranging from 0.10 to 0.24 usually places an individual in the "pre-frail" category, which indicates a moderate risk level and represents an important opportunity for early intervention and preventative strategies.

The distribution of the frailty index across different age groups enables researchers and clinicians to determine the proportion of subjects who are considered frail or pre-frail within each group. This stratification provides invaluable insights into how frailty prevalence changes with age, assisting in the allocation of healthcare resources and the design of targeted interventions. Such measures aim to mitigate the progression of frailty, thereby enhancing health outcomes and quality of life in the aging population.

Several studies have characterised the relationship between Frailty Index and the prevalence of frailty at a given age.

Rockwood et al. characterized the distribution of frailty index in human and mice as a function of age and built an association between incremental Frailty Index scores and a hazard ratio for all-cause mortality\cite{Rockwood2017}. Rockwood and Howlett later showed that the Frailty Index distribution fits a gamma distribution for younger ages, with a long right tail, indicating some younger people have many health problems. However, as people age, the distribution becomes normal, showing how health problems are not just for a few, but affect everyone\cite{Rockwood2019}. These observations were confirmed by a longitudinal study in Korea, characterizing the evolution of the distributions of Frailty Index as a function of age and showed that the median Frailty Index is approximately 0.5 among females aged 95-98\cite{Baek2022}. In determining the proportion of frail people at a given age, we calculate the proportion of subjects above the 0.25 Frailty Index threshold, as per definition. We perform a normal approximation of the gamma distribution and calculate the proportion of frail subjects at each age as

\begin{equation}
\pi_{x}^{frail} = 1 - \Phi\left( \frac{0.25\  - \ {FI}_{x}}{\sigma_{x}^{FI}} \right)
\end{equation}

Being $\pi_{x}^{frail}$ the proportion of frail subject at age x, ${FI}_{x}$ the Frailty Index at age x, $\sigma_{x}^{FI}$ the standard deviation of the Frailty Index at age x and $\Phi(.)$ the cumulative distribution function of a standard normal distribution.

\begin{figure}[h]
\centering
\includegraphics[width=\textwidth]{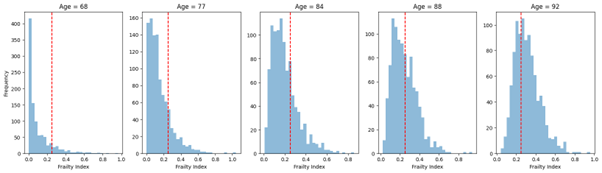}
\caption{Distributions of Frailty Index as age gets larger (left to right). The vertical dashed red line represents the 0.25 Frailty Index threshold, above which a person is classified as frail.}
\label{fig:frailty_distribution}
\end{figure}

\subsection{Model Development}

In this section, we explain the data sources and methods used to meet the goals of the manuscript. As we mentioned, one main challenge of using Frailty as an endpoint is that it occurs relatively late in life, with trials like the DO-HEALTH trial enrolling subjects older than 65 (70) years old\cite{McNeil2018}. Therefore, finding studies that target individuals younger than 65 is crucial. Several studies, including the CALERIE trial reported the effects of dietary interventions on younger subjects (less than 50 years old)\cite{Waziry2023}. The work by Mak et al.\cite{Mak2023} modelled the evolution of \textit{Frailty Index} as a function of the evolution of five epigenetic clocks, including clocks that assess the current estimate of biological age (PCHorvathAge, PCHannumAge, PCPhenoAge, PCGrimAge, DunedinPACE) and clocks that assess the speed or pace of ageing, e.g. the DunedinPACE clock. The study focuses on subjects older than 50 years old, thus enabling the generation of evidence for more than an additional decade compared to the early studies (for example\cite{McNeil2018}). From a modelling standpoints, the study used a Dual-Change-Score-Model\cite{Wiedemann2022} to model the dependencies between Frailty Index and the epigenetic clocks. Importantly, the study demonstrated how changes in the proportion of frail subjects in a population, as affected by the value of recorded Frailty Indices, are influenced by the current proportion of frail individuals and by the value of a specific epigenetic clock.

Their study concluded that there is no statistically significant relationship between any of the clocks measuring the estimate of biological age, but there is a dependency between Frailty and DunedinPACE. The authors estimated the parameters affecting the relationship between DunedinPACE and Frailty Index, with a one-directional coupling effect DunedinPACE → Frailty (reported in Table~\ref{tab:parameters}).

The structure of the model identified by Mak and coworkers is described by the Equation (\ref{eq:delta_fi})

\begin{equation}\label{eq:delta_fi}
{\mathrm{\Delta}FI}_{t} = \ \alpha \cdot {FI}_{S} + \beta_{FI} \cdot {FI}_{t - 1} + \gamma_{Clock \rightarrow FI} \cdot {Clock}_{t - 1}
\end{equation}

Where a 2-year cycle is assumed (thus the time gap between two consecutive cycles t-1 and t is 2 years) and the parameters and their descriptions are outlined in Table~\ref{tab:parameters}.

According to the modeling assumptions of Mak and colleagues, the change in DunedinPACE over time is a linear function of age, which can be described by the equation below:

\begin{equation}\label{eq:delta_clock}
{\mathrm{\Delta}Clock}_{t} = \ \beta_{Clock} + {Clock}_{t - 1}
\end{equation}

We extend the quantitative model by Mak and colleagues to determine how changes in the profile of the DunedinPACE Clock affect changes in the Frailty Index over a 40-year time period, namely between 50 and 90 years of age.
\begin{table}[ht]
\centering
\small
\setlength{\tabcolsep}{4pt}
\begin{tabular}{lp{5.5cm}l}
\toprule
\textbf{Parameter} & \textbf{Description} & \textbf{Value} \\
\midrule
$\alpha$ & & \\
$FI_{50}$ & Frailty Index at 50 years & 6.55 (\%) \\
$FI_{50}^{gender}$ & Effect of gender (female vs male) on $FI_{50}$ & \\
$FI_{S}$ & Slope on frailty & -12.31 \\
$FI_{S}^{gender}$ & Effect of gender (female vs male) on $FI_{S}$ & \\
$\beta_{FI}$ & Coefficient on FI & \\
$\gamma_{Clock \rightarrow FI}$ & Coupling coefficient between Clock (DunedinPACE) and FI & 1.19 \\
$Clock_{50}$ & 10x value of DunedinPACE at age 50 & 10.04 \\
$\beta_{Clock}$ & & 0.06 \\
$HR_{\frac{frail}{robust}}$ & Hazard ratio in all-cause mortality between frail and robust people & 2.40 \\
\bottomrule
\end{tabular}
\caption{Parameters employed in the model with corresponding values}
\label{tab:parameters}
\end{table}

\begin{figure}[ht]
\centering
\includegraphics[width=\textwidth]{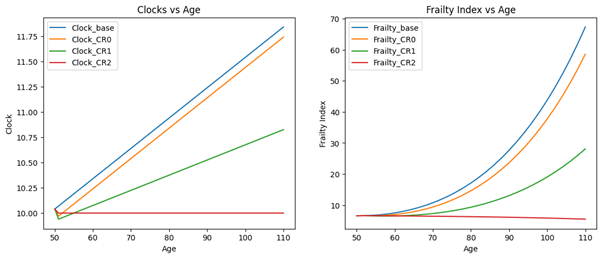}
\caption{Overview of the evolution of the DunedinPACE clock (left panel) and Frailty Index (right panel) as a function of Age. The blue line corresponds to the baseline evolution of DunedinPACE as a function of age. The orange line corresponds to the putative evolution with changes in DunedinPACE only governed by the findings of the CALERIE trial, i.e. a shift of 0.01 applied to the y-axis only [CR0]. The green line assumes that changes in diet and lifestyle govern both the offset on the y-axis and the slope of the frailly line as a function of age [CR1], while the red line corresponds to a scenario in which individuals manage to keep their DunedinPACE clock value constantly at 1.0 [CR2], thereby leading to a favorable evolution of Frailty Index.}
\label{fig:clocks_frailty}
\end{figure}

We construct three different types of evolution of the DunedinPACE clock, which depend on assumptions on the evolution of the evidence base. The three scenarios are:

\begin{enumerate}
    \item Interventions, such as caloric restrictions (as from the CALERIE study) lead to a reduction of DunedinPACE of 0.01 with respect to the expected DunedinPACE evolution (cite CALERIE) as a function of age. This can be seen as a simple shift of the DunedinPACE evolution along the y-axis. (Figure~\ref{fig:clocks_frailty}) [CR0]. 
    
    \item In addition to a shift on the y-axis, interventions affect the slope of the natural evolution of DunedinPACE, thus reducing the rate of growth of the speed of aging year-on-year [CR1].
    
    \item A third scenario is based on the identification of a spectrum of interventions, which will set the pace of aging exactly at 1.0 (the natural speed of aging) or slightly below [CR2]. 
\end{enumerate}

The three scenarios outlined above are linked to the evolution of \textit{Frailty Index} as depicted in Figure~\ref{fig:clocks_frailty}.

\begin{figure}[ht]
\centering
\includegraphics[width=\textwidth]{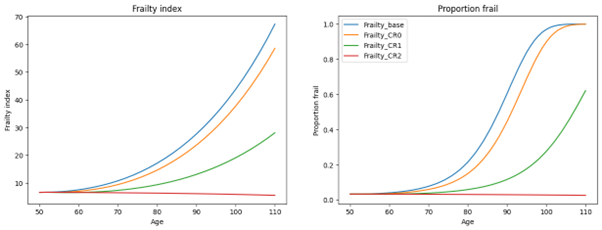}
\caption{Overview of the evolution of the proportion of frail subjects (right panel) as a function of Frailty Index (left panel). The blue line present the base case for the evolution observed in the dual-change score model\cite{Mak2023}. The orange line to the CR0 scenario, in which a reduction in the DunedinPACE is applied on the y-axis only. The green line assumes that changes in diet and lifestyle govern both the offset on the y-axis and the slope of the frailly line as a function of age [CR1], while the red line corresponds to a scenario in which individuals manage to keep their DunedinPACE clock value constantly at 1.0 [CR2].}
\label{fig:frailty_proportion}
\end{figure}

As outlined in Section 2.2, the link between Frailty Index and the prevalence of frailty is governed by Equation (1). After calibrating the value of the standard deviation $\sigma_{x}^{FI}$ to match the curves representing the proportion of frailty as a function of age observed in a Dutch study\cite{Hoogendijk2019}, we obtain the curves as in Figure~\ref{fig:frailty_proportion}.

\subsection{Estimation of the lifespan changes depending on Frailty}

We use mortality data from selected countries (e.g. Switzerland)\cite{Ruckstuhl2023} to measure how the survival function changes according to changes in the fraction of frail people in the general population. We construct two survival curves $S_{x}^{robust}$ and $S_{x}^{frail}$, which show the probability of survival of robust and frail people respectively. To connect the robust and frail populations, we use the relationship between the force of mortality $\mu_{x}$ for the two populations. The relationship between the force of mortality, or hazard, of the robust and the frail population has been studied extensively in recent years. Peng et al. did a systematic review that showed the risks of all-cause and cause-specific mortality in frail and robust people living in the community. The authors found a hazard ratio of 2.40, which represents the higher risk of frail people dying from any cause compared to robust people\cite{Peng2022}. This is described by the equation below

\begin{equation}\label{eq:hazard_ratio}
\frac{\mu_{x}^{frail}}{\mu_{x}^{robust}} = {HR}_{\frac{frail}{robust}} = 2.40\ 
\end{equation}

Where $\mu_{x}^{frail}$ and $\mu_{x}^{robust}$ represent the hazard functions at age $x$ for the frail and the robust populations respectively. Starting off from the general population mortality, we construct new survival curves that separate out the robust and frail individuals from the general population. Our assumption here is that the hazard ratio of frail vs non-frail is constant over time and does not depend on age. In this fashion, we obtain two distinct survival curves, one representing the population as if all the individuals were frail and one representing the population as if everyone was robust. The standard relationship linking Survival and hazard is well known\cite{Kleinbaum2012}

\begin{equation}\label{eq:survival_hazard}
\mu_{x} = - \log\left( \frac{S_{x + 1}}{S_{x}} \right)
\end{equation}

In our work we relied on this discrete approximation of the number of death events at each year and built survival curves that capture the contribution of both the frail and the non-frail populations. This is achieved by building a survival curve including both frail and non-frail as

\begin{equation}\label{eq:survival_total}
{S_{x}^{total} = \pi_{x}^{frail}S}_{x}^{frail}{+ \left( 1 - \pi_{x}^{frail} \right)S}_{x}^{robust}
\end{equation}

Where $\pi_{x}^{frail}$ is the proportion of frail people at age $x$, which changes according to the scenarios outlined in Figure~\ref{fig:frailty_proportion}. Since our analysis focuses on subjects older than 50 years old, we have $S_{50}^{robust} = S_{50}^{frail} = S_{50}^{total} = 1$ (in standard mortality tables this value is actually 1e5). Let $D_{x}$ be the number of deaths at age $x$. A proportion of these deaths is attributable to frailty and a proportion will be among robust people. By factoring in the evolution of the proportion of frail people over time $\pi_{x}^{frail}$ in Equation~\eqref{eq:survival_hazard} and accounting for the Hazard Ratio outlined in Equation~\eqref{eq:hazard_ratio}, we can build independent survival curves for the frail and non-frail populations (see Results section).

We have now shown how changes in epigenetic clocks, such as DunedinPACE, can affect changes in frailty, which in turn affect changes in the expected survival or lifespan. We will investigate how changes in DunedinPACE can be achieved with caloric restriction, as demonstrated by the CALERIE trial, which had a relatively short follow up time of 2 years.

\subsection{Value and savings to Healthcare systems}

In order to demonstrate the effects of changes in Frailty in the overall healthcare system costs, we relied on a study comparing the costs born to selected healthcare systems in frail vs. robust subjects, addressing frailty as an overall health burden\cite{Hoogendijk2019}. Notably, the authors outline an increase in spending in Frail subjects compared to robust subjects. The frailty instruments used were not consistent among the various resources utilized. In a Study carried out in the Netherlands, the authors used the Groningen Frailty Indicator and used healthcare insurance data to show an increase in mean annual costs of approximately EUR 15000 in frail subjects compared to non-frail subjects aged 65 and above\cite{Peters2015}. In the USA, hospital registration data showed additional costs in the 6 months following colorectal surgery slightly below USD 50000 if frail subjects vs robust subjects aged 65 and above\cite{Robinson2011}. A minor difference was observed in a study which gathered data from a government database in Australia. Among patients with a mean age of 79.5 years, frail patients showed mean total health-care costs increase in the 6 months post hospital admission approximately AUD 9000 higher compared to non-frail patients\cite{Comans2016}.

Costs have been converted to 2024 costs in Switzerland (CHF) based on the cumulative inflation since the year of observation (Year column - Table~\ref{tab:costs}) and 2024, the currency conversion rate and an adjustment for the cost of living between Switzerland and the listed countries. The inflation rate for Australia has been extracted from the Australian Bureau of Statistics\cite{ABS2023}. The inflation rate for the USA has been extracted from the US Bureau of Labor Statistics\cite{BLS2024} and the inflation rate for the Netherlands has been extracted from Statistics Netherlands (CBS)\cite{CBS2024}. The purchasing parity conversion was extracted from the OECD Purchasing Power Parities for GDP and related indicators table\cite{OECD}. The exchange rate of CHF to AUD and USD have been sourced from the Swiss National Bank\cite{SNB}.

\begin{table}[ht]
\centering
\small
\setlength{\tabcolsep}{3pt}
\begin{tabular}{lrrrcc}
\toprule
\textbf{Country} & \textbf{Year} & \multicolumn{2}{c}{\textbf{Annual Cost}} & \multicolumn{2}{c}{\textbf{2024 CHF Adjusted}} \\
\cmidrule(lr){3-4}\cmidrule(lr){5-6}
 & & \textbf{Frail} & \textbf{Non-frail} & \textbf{Frail} & \textbf{Non-frail} \\
\midrule
Netherlands & 2015 & EUR 30,792 & EUR 15,611 & 52,168 & 26,448 \\
Australia & 2016 & AUD 28,906 & AUD 19,905 & & \\
USA & 2011 & USD 76,363 & USD 27,731 & & \\
\bottomrule
\end{tabular}
\caption{Healthcare costs of frail vs non-frail individuals across countries with adjustments to 2024 Swiss costs (CHF) accounting for inflation and purchasing power.}
\label{tab:costs}
\end{table}

While the study in the Netherlands shows a direct impact on total annual healthcare costs, which can be directly translated to potential savings for decision-makers, the studies in the US and Australia are focused on post-operative costs for selected categories. These costs are difficult to translate into potential direct savings for a healthcare system. A study in Switzerland calculated the proportion of people admitted to hospital according to frail status, showing that the incidence of hospital admission over a 10-year period is approximately 6-fold higher in frail or pre-frail people compared to robust individuals\cite{Luini2023}. Nonetheless, since the specific focus of this study is on a population admitted to hospital, we believe its findings are not applicable to our context.

\subsection{Value to individuals}

In this section, we outline the economic framework to value improvements in lifespan and health span, as outlined in the work by Murphy and Topel\cite{Murphy2006} and Scott et al.\cite{Scott2023a,Scott2023b}. The work defines the lifetime utility for an individual of age $a$, defined as

\begin{equation}\label{eq:lifetime_utility}
utility(a) = \int_{a}^{\infty}{H(t)u\left( c(t),l(t) \right)\widetilde{S}}(t,a)e^{- \rho(t - a)}dt
\end{equation}

Were the elements of the Equation~\eqref{eq:lifetime_utility} being:

\begin{itemize}
    \item $H(t)$: intangible value of health, which affects the quality of life (health-related) without affecting mortality
    \item $u(c(t), l(t))$: utility to an individual, which in turn depends on
    \item $c(t)$: consumption, or the ability of an individual to purchase good or consume services
    \item $l(t)$: non-market or leisure time, that is the number of hours in a year an individual does not need to work and can enjoy their free time
    \item $\widetilde{S}(t,a)$: the probability of survival from age $a$ to $t$
    \item $\rho$: the discount rate or the rate of time preference
\end{itemize}

Under a perfect annuity market, the expected discounted value of future consumption corresponds to the expected wealth\cite{Murphy2006}, that is:

\begin{equation}\label{eq:expected_wealth}
A(a) + \int_{a}^{\infty}{\left\lbrack y(t) - c(t) \right\rbrack\widetilde{S}(t,a)e^{- r(t - a)}dt = 0}
\end{equation}

In the equation the following terms appear:
\begin{itemize}
    \item $A(a)$: the initial assets, or the wealth at age $a$
    \item $r$: the interest rate
    \item $y(t)$: the life-contingent income
\end{itemize}

We note that the life-contingent income $y(t)$ is:

\begin{equation}\label{eq:income}
y(t) = w(t)\left\lbrack 1 - l(t) \right\rbrack + b(t)
\end{equation}

Where $w(t)$ is the wage per time period, $l(t)$ is the proportion of leisure time and $b(t)$ is a non-wage income, such as a pension, that is received without working-hours associated to generating it.

The key idea behind determining the value of a year of life to individual is to maximize the utility associated to the remainder of one's life, subject to the constraints due to the disposable capital, as the sum of the currently available assets and the future income. The value of life is calculated as the marginal rate of substitution between the instantaneous mortality hazard at age $a$, and the assets accumulated $A(a)$.

The corresponding value of residual life at age $a$ is then calculated as

\begin{equation}\label{eq:residual_life}
V(a) = \ \int_{a}^{\infty}{v(t)}\widetilde{S}(t,a)e^{- r(t - a)}dt
\end{equation}

Which is independent on the rate of time preference $\rho$ and where $v(t)$ is the value of a life year at age $t$, larger or equal than $a$. The value of a life year is in turn a function of the wage $w(t)$, the total working hours, leisure time and the marginal utility of consumption\cite{Murphy2006,Scott2021}.

Under this construct, the willingness to pay for changes in technological advancements $\xi$, such as those coming from the information coming from epigenetic clocks measuring the speed of aging -- in turn inducing lifestyle changes, or pharmaceutical that target the ageing process is

\begin{equation}\label{eq:wtp}
{WTP}_{\xi}(a) = \underbrace{\int_{a}^{\infty}{v(t)}\mathrm{\Delta}_{\xi}S^{*}(t,a)dt}_{\text{lifespan}} + \underbrace{\int_{a}^{\infty}{\frac{\mathrm{\Delta}_{\xi}H(t)}{H(t)}\frac{u(c(t),l(t))}{u_{c}(c(t),l(t))}}dt}_{\text{healthspan}}
\end{equation}

Where $S^{*}$ is the discounted survival, $\mathrm{\Delta}_{\xi}S^{*}$ and $\mathrm{\Delta}_{\xi}H$ denote the change in Survival and Health induced by the technological advancement $\xi$ and $u_{c}$ is the marginal utility of consumption. The expression contains independent terms that represent contributions to the willingness to pay with respect to changes in lifespan and healthspan.

\subsection{Linking Frailty to Health}

The above-mentioned studies by Scott and Topel presented changes in health as a function of changes in frailty via disability. In our work, we modeled health as a linear function of Frailty Index, as observed in previous studies that outlined the inverse relationship between Frailty Index and health related quality of life utility\cite{Nathiya2023}. We simplify these observations further and model health as $H(t) = 1 - FI(t)$.

We measure the outcomes of the model in terms of the number and proportion of individuals who are frail, have chronic diseases, or die at each age.

\section{Results}

\subsection{Healthcare System Cost Savings}

We quantified the potential long-term savings to healthcare systems based on our modeled intervention strategies derived from DunedinPACE data, comparing these savings against current projections of aging-related healthcare expenditures.

Our analysis revealed significant savings across all intervention scenarios compared to the baseline control (no aging-pace intervention). We constructed two distinct scenarios: Scenario 1 examined a hypothetical cohort followed from age 50 to 90, assuming all individuals remained alive throughout the 40-year period; Scenario 2 incorporated natural mortality patterns, progressively reducing cohort size over time. Notably, our analysis excluded costs associated with the health technology enabling DunedinPACE monitoring and control.

\begin{figure}[ht]
\centering
\small
\includegraphics[width=\textwidth]{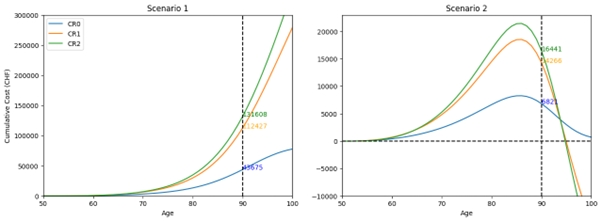}
\caption{Cumulative healthcare system costs under two scenarios. Scenario 1 follows a cohort from age 50 to 90 assuming no mortality. Scenario 2 incorporates natural mortality progressively reducing cohort size. Lines show cumulative savings under three epigenetic clock control strategies: CR0 (blue), CR1 (orange), and CR2 (green).}
\label{fig:savings}
\end{figure}

The savings became increasingly pronounced over time as intervention effects accumulated, delaying frailty onset and progression. Under Scenario 1 assumptions, cumulative savings reached CHF 43,675 with the CR0 DunedinPACE control strategy, CHF 112,427 with CR1, and CHF 131,608 with CR2 (Figure~\ref{fig:savings}, left panel).

Scenario 2 demonstrated more complex dynamics. As frailty-related mortality increased with age, robust individuals comprised an increasing proportion of survivors, particularly in intervention scenarios. This created the cumulative savings pattern shown in Figure~\ref{fig:savings} (right panel). Savings peaked at approximately age 85 (35 years post-intervention initiation), with CR2 showing the most pronounced benefits compared to CR1 and CR0. Interestingly, the trend reversed after age 95, eventually resulting in net losses to healthcare systems. This counterintuitive finding reflects the intervention's success in extending lifespan, thereby increasing the proportion of individuals requiring late-life healthcare services—a phenomenon observed in other therapeutic domains, such as oncology, where survival extension often necessitates continued healthcare expenditures.

\subsection{Value to Individuals}

While healthcare system savings provide valuable perspective on intervention cost-effectiveness from a payer standpoint, this analysis has limitations. It focuses solely on frailty-related direct costs, omitting costs indirectly linked to frailty (e.g., comorbidities such as dementia, cancer, or cardiovascular disease). Additionally, it extrapolates Dutch healthcare data to the Swiss context, where frailty-related expenditures may differ significantly.

We therefore extended our analysis to quantify individual value, building on the economic framework outlined in the Methods section. We determined lifespan and healthspan changes induced by the three CR regimens, combining these with life-year values $v(t)$ and marginal consumption utility to generate willingness-to-pay curves for the health technology $\xi$.

\begin{figure}[ht]
\centering
\small
\includegraphics[width=\textwidth]{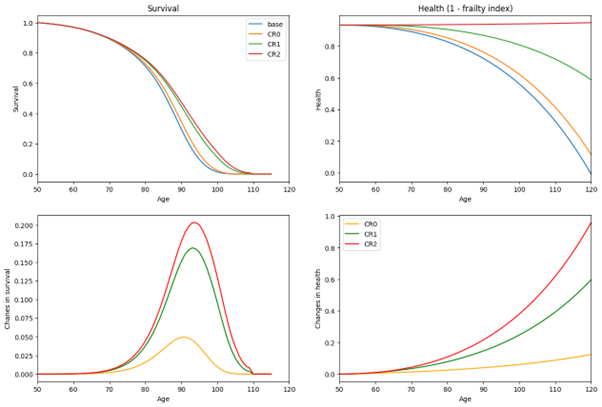}
\caption{Lifespan (survival, left column) and healthspan (right column) evolution under three control scenarios. Top row shows baseline evolution (blue) and corresponding curves for CR0 (orange), CR1 (green), and CR2 (red). Bottom row shows changes in survival and health for each control scenario relative to baseline.}
\label{fig:lifespan_healthspan}
\end{figure}

Survival curves shifted rightward with increasing DunedinPACE control, with effects most pronounced in individuals over 80—those most likely to be frail in the baseline scenario (Figure~\ref{fig:lifespan_healthspan}, top-left panel). The effect magnitude peaked around age 90 (Figure~\ref{fig:lifespan_healthspan}, bottom-left panel). Regarding healthspan, control regimens significantly reduced health-related quality-of-life decline, maintaining high quality-of-life levels throughout aging, with the CR2 scenario even showing slight improvements (Figure~\ref{fig:lifespan_healthspan}, right panels).

Mean life expectancy was 84.34 years in the baseline scenario, increasing to 85.11 years with CR0, 87.22 years with CR1, and 87.92 years with CR2—representing life expectancy gains ranging from 0.77 to 3.58 years (Table~\ref{tab:lifeexp}).

\begin{table}[ht]
\centering
\small
\setlength{\tabcolsep}{5pt}
\begin{tabular}{lrr}
\toprule
\textbf{Scenario} & \textbf{Life expectancy} & \textbf{Life expectancy gain} \\
\midrule
Base & 84.34 & 0.00 \\
CR0 & 85.11 & 0.77 \\
CR1 & 87.22 & 2.89 \\
CR2 & 87.92 & 3.58 \\
\bottomrule
\end{tabular}
\caption{Life expectancy across scenarios and gains relative to baseline.}
\label{tab:lifeexp}
\end{table}

\begin{figure}[ht]
\centering
\small
\includegraphics[width=\textwidth]{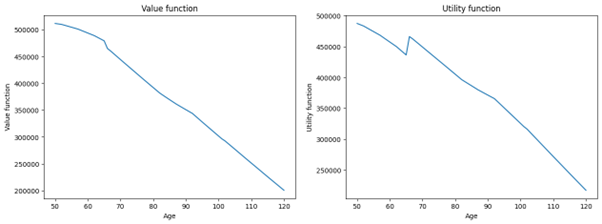}
\caption{Value function $v(t)$ (left) and marginal utility function for consumption (right) across age.}
\label{fig:value_utility}
\end{figure}

Calculating willingness-to-pay for lifespan and healthspan changes required incorporating profiles of $v(t)$ and marginal consumption utility, calibrated using parameters from Scott et al.~\cite{Scott2023b}. The monetary value $v(t)$ showed a decreasing trend with age (Figure~\ref{fig:value_utility}, left panel), while the marginal consumption utility exhibited a more complex pattern (Figure~\ref{fig:value_utility}, right panel). Notable inflection points occur at age 65 due to retirement: $v(t)$ drops reflecting reduced income, while utility increases sharply, reflecting increased leisure time availability.

\begin{figure}[ht]
\centering
\small
\includegraphics[width=\textwidth]{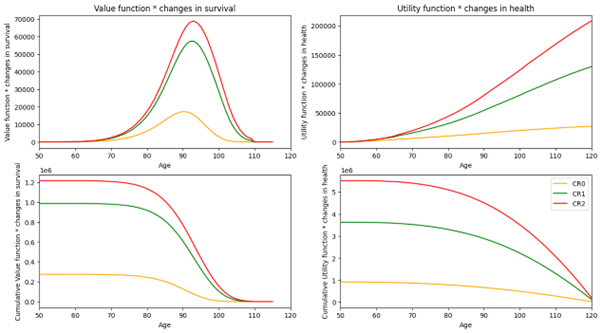}
\caption{Annual willingness-to-pay for additional life based on survival changes (top-left) and cumulative willingness-to-pay from age 50 (bottom-left). Annual utility values from health improvements (top-right) and corresponding cumulative utility from age 50 (bottom-right).}
\label{fig:wtp}
\end{figure}

The CR2 scenario generated the highest willingness-to-pay due to its impact on both lifespan and life quality. Lifestyle changes affecting DunedinPACE yielded highest willingness-to-pay values between ages 80-100, when frailty reduction benefits are most pronounced (Figure~\ref{fig:wtp}, left panels). Health improvements affected willingness-to-pay from earlier ages, with notable increases around age 65 due to retirement-related leisure time valuation (Figure~\ref{fig:wtp}, right panels).

Cumulative willingness-to-pay curves demonstrated substantial valuations for both survival and health changes, exceeding CHF 1.2 million for survival and CHF 5.5 million for health improvements under the strictest DunedinPACE control scenario CR2 (Table~\ref{tab:wtp}). Note that these values are undiscounted, assuming no personal time preference.

\begin{table}[ht]
\centering
\small
\setlength{\tabcolsep}{4pt}
\begin{tabular}{lrrr}
\toprule
\textbf{Scenario} & \textbf{Survival Changes} & \textbf{Health Changes} & \textbf{Total Value} \\
\midrule
CR0 & 274.68 & 916.92 & 1,191.59 \\
CR1 & 986.78 & 3,628.88 & 4,615.66 \\
CR2 & 1,216.19 & 5,515.61 & 6,731.80 \\
\bottomrule
\end{tabular}
\caption{Willingness-to-pay at age 50 for changes in survival, health, and their combination over remaining lifespan (thousands of CHF).}
\label{tab:wtp}
\end{table}

\subsection{Squaring the Survival Curve}

"Squaring the survival curve" is a concept in public health and gerontology describing an ideal survival trajectory where individuals remain healthy until very near their life expectancy, followed by a rapid decline~\cite{Caselli2021}. This contrasts with traditional survival curves showing steady decline from birth with accelerating mortality at advanced ages.

The squared curve represents a compression of morbidity scenario—individuals live full, healthy lives with minimal disability or chronic disease until shortly before death, rather than experiencing prolonged periods of diminished health.

\begin{figure}[ht]
\centering
\small
\includegraphics[width=\textwidth]{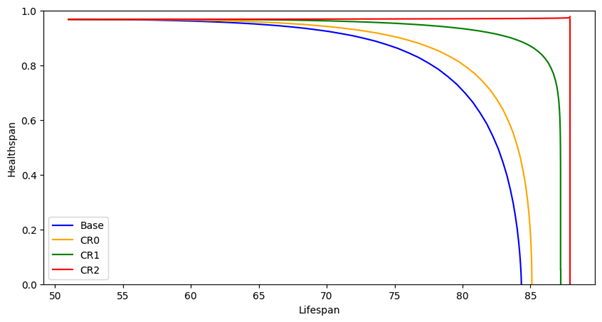}
\caption{Squaring the survival curve: healthspan versus lifespan for CR0 (orange), CR1 (green), and CR2 (red) scenarios. Stricter DunedinPACE control produces increasingly rectangular survival curves, indicating compressed morbidity.}
\label{fig:squaring}
\end{figure}

Our assessment of DunedinPACE control scenarios demonstrated progressively more rectangular health-versus-survival curves with stricter control regimens. The CR2 strategy in particular produced a more squared curve, suggesting successful morbidity compression and the potential for individuals to maintain high quality of life until near death (Figure~\ref{fig:squaring}).

\subsection{Quality-Adjusted Life Year Gains}

Quality-Adjusted Life Years (QALYs) combine quantity and quality of life into a single metric used in health economics to assess intervention value~\cite{Briggs2023}. One QALY represents one year in perfect health.

The Incremental Cost-Effectiveness Ratio (ICER) evaluates new medical interventions by dividing cost difference by QALY difference:

\begin{equation}\label{eq:icer}
ICER = \ \frac{\mathrm{\Delta}_{\xi}Cost}{\mathrm{\Delta}_{\xi}QALY}
\end{equation}

where $\mathrm{\Delta}_{\xi}QALY$ and $\mathrm{\Delta}_{\xi}Cost$ represent QALY and cost changes attributable to health technology $\xi$. Our analysis to this point excluded intervention costs, so ICER calculations will inform pricing strategies for the technology.

ICER thresholds represent maximum willingness-to-pay for one additional QALY. The World Health Organization suggests thresholds of 1-3× GDP per capita~\cite{Woods2016}. With Switzerland's GDP per capita approximating CHF 100,000 (2023-2024)~\cite{IMF}, relevant ICER thresholds range from CHF 100,000-300,000.

Our model calculated QALY gains of 1.07 for CR0, 3.90 for CR1, and 5.32 for CR2. Using these thresholds, willingness-to-pay ranges from approximately CHF 107,000 (CR0, lowest threshold) to over CHF 1.5 million (CR2, highest threshold) (Table~\ref{tab:qaly}).

\begin{table}[ht]
\centering
\small
\setlength{\tabcolsep}{3pt}
\begin{tabular}{lrrrr}
\toprule
\textbf{Scenario} & \textbf{QALY} & \textbf{$\mathbf{\Delta}$QALY} & \textbf{ICER} & \textbf{WTP} \\
\midrule
Base & 80.19 & 0.00 & 100,000 & 0 \\
CR0 & 81.26 & 1.07 & 100,000 & 107,173 \\
CR1 & 84.09 & 3.90 & 100,000 & 389,901 \\
CR2 & 85.51 & 5.32 & 100,000 & 531,628 \\
\midrule
Base & 80.19 & 0.00 & 200,000 & 0 \\
CR0 & 81.26 & 1.07 & 200,000 & 214,346 \\
CR1 & 84.09 & 3.90 & 200,000 & 779,803 \\
CR2 & 85.51 & 5.32 & 200,000 & 1,063,257 \\
\midrule
Base & 80.19 & 0.00 & 300,000 & 0 \\
CR0 & 81.26 & 1.07 & 300,000 & 321,519 \\
CR1 & 84.09 & 3.90 & 300,000 & 1,169,704 \\
CR2 & 85.51 & 5.32 & 300,000 & 1,594,885 \\
\bottomrule
\end{tabular}
\caption{QALY gains across intervention scenarios at three ICER threshold levels (CHF 100,000-300,000).}
\label{tab:qaly}
\end{table}

The expected outcomes of the proposed DunedinPACE monitoring and intervention strategy include reduced frailty and chronic disease burden, improved individual health and wellbeing across the lifespan, enhanced healthcare system efficiency, and positive economic and social welfare effects.

Preliminary evidence from our pilot study with 100 Dunedin Study participants demonstrates the feasibility and acceptability of DunedinPACE assessments and personalized interventions. Participants reported high satisfaction and adherence, with improved biomarkers and health outcomes after six months of follow-up, supporting the viability of the proposed approach.

\section{Discussion}

In this section, we discuss the implications of our model and forecast results, address methodological strengths and limitations, outline challenges to implementation, and propose directions for future research.

\subsection{Summary of Key Findings}

We have developed a novel biomarker-based monitoring and intervention strategy for longevity medicine based on routine DunedinPACE assessments in clinical settings. Our economic modeling demonstrates that DunedinPACE-based intervention scenarios yield significant benefits compared to baseline projections across three domains:

Healthcare systems could realize substantial cost savings through delayed frailty onset and progression, with cumulative savings reaching CHF 131,608 per individual under the CR2 scenario in our 40-year cohort model. These savings exhibit temporal dynamics, peaking around age 85 before potentially reversing due to extended lifespan and associated late-life care needs.

Individual value assessments revealed considerable willingness-to-pay for interventions affecting aging pace, ranging from CHF 1.2 million for survival improvements to CHF 5.5 million for health-related quality-of-life enhancements under the CR2 scenario, suggesting strong economic incentives for personal investment in aging-moderation strategies.

Our QALY analysis showed meaningful gains across intervention scenarios (1.07-5.32 additional QALYs), translating to substantial value across accepted willingness-to-pay thresholds (CHF 100,000-300,000 per QALY).

The main implication of our findings is that epigenetic pace-of-aging assessments could provide a valuable framework for identifying accelerated aging trajectories and implementing targeted interventions, potentially yielding significant economic and health benefits at both individual and societal levels.

\subsection{Methodological Strengths}

Our approach leverages the DunedinPACE clock, a robust biomarker capturing epigenetic change rates with demonstrated predictive capacity for multiple age-related outcomes. This clock provides a dynamic assessment of biological aging processes that reflects past exposures, current status, and future health trajectory.

The incorporation of the Frailty Index as our primary clinical outcome measure allows for translation between molecular-level epigenetic changes and clinically relevant functional status, providing a coherent pathway from biological processes to healthcare utilization patterns. By linking these elements to economic metrics including healthcare costs, willingness-to-pay, and QALYs, we establish a comprehensive framework for evaluating aging-focused interventions.

Furthermore, our analysis begins at age 50, approximately 15 years earlier than the inclusion criteria for most aging-targeted trials, potentially expanding the intervention window during which aging modification strategies could yield meaningful benefits.

\subsection{Limitations and Constraints}

Several important limitations must be considered when interpreting our findings. First, our model relies on connecting results from disparate cohort studies. The dual change score model described by Mak et al.~\cite{Mak2023} connecting DunedinPACE to Frailty Index progression was developed in a cohort distinct from the CALERIE trial that demonstrated caloric restriction effects on DunedinPACE~\cite{Waziry2023}. This cross-cohort extrapolation introduces uncertainty regarding generalizability.

Second, our intervention scenario focuses exclusively on caloric restriction as the primary mechanism for modifying DunedinPACE trajectories. While evidence supports this connection, it represents only one of many potential intervention pathways. Our model does not account for other lifestyle, pharmaceutical, or combination approaches that might influence aging pace through different mechanisms.

Third, our analysis examines frailty as the primary clinical outcome without explicitly modeling specific disease trajectories. This approach, while valuable for overall healthcare utilization forecasting, does not capture the heterogeneity of age-related pathologies across organ systems. Future research should expand beyond frailty to explicitly model how DunedinPACE and aging interventions affect specific age-related conditions, particularly neurological disorders, cardiovascular disease, and musculoskeletal conditions.

Fourth, our economic model necessarily simplifies the complex healthcare landscape, focusing on direct frailty-related costs while potentially omitting indirect expenditures. Additionally, transportation of Dutch healthcare cost data to the Swiss context introduces potential inaccuracies in absolute savings estimates.

Finally, our model assumes consistent intervention adherence and efficacy across the lifespan, which may not reflect real-world implementation challenges where intervention compliance may fluctuate and effectiveness might diminish over time or vary across individuals.

\subsection{Future Research Directions}

Based on our findings and acknowledged limitations, several priorities emerge for future research:

\begin{itemize}
    \item \textbf{Expanded disease modeling:} Future studies should extend beyond frailty to explicitly model how DunedinPACE and aging interventions affect specific age-related conditions, particularly neurological disorders, cardiovascular disease, and musculoskeletal conditions. This would provide more granular understanding of intervention benefits across diverse pathophysiological processes.
    
    \item \textbf{Intervention diversity:} Research should explore multiple intervention modalities beyond caloric restriction, including exercise regimens, sleep optimization, stress reduction, pharmaceutical agents (such as metformin, rapamycin), and combination approaches, evaluating their relative and synergistic effects on aging biomarkers.
    
    \item \textbf{Cohort validation:} Validation studies should assess the DunedinPACE-frailty relationship across diverse populations, including different ethnicities, socioeconomic backgrounds, and geographical regions to establish generalizability and identify potential subgroup variations.
    
    \item \textbf{Implementation science:} Studies examining real-world implementation challenges, including intervention adherence, provider adoption barriers, healthcare system integration, and cost-effective delivery models, will be essential for translating theoretical benefits into practical outcomes.
    
    \item \textbf{Longitudinal validation:} Long-term prospective studies following individuals from middle age through late life with regular DunedinPACE assessments and comprehensive health outcomes tracking would provide more definitive evidence regarding predictive validity and intervention efficacy.
\end{itemize}

\subsection{Ethical and Practical Considerations}

The implementation of epigenetic clock assessments and personalized aging interventions raises important ethical considerations. Privacy and consent frameworks must be established to govern the collection, storage, and utilization of epigenetic data. Protections against potential discrimination based on biological age assessments in insurance, employment, or healthcare access contexts will be essential.

Equity concerns also warrant attention, as access to advanced testing and intervention resources could exacerbate existing health disparities if not thoughtfully distributed. Healthcare systems must consider how to integrate such technologies in ways that reduce rather than amplify inequalities.

From a practical perspective, implementation barriers include limited awareness among healthcare providers regarding epigenetic aging concepts, insufficient technical infrastructure for routine epigenetic assessments, and absence of established clinical guidelines for interpreting and acting upon DunedinPACE results. Coordinated efforts across research, clinical, policy, and industry domains will be necessary to overcome these challenges.

\section{Conclusion}

In conclusion, we have developed a novel biomarker-based monitoring and intervention strategy for longevity medicine based on routine DunedinPACE assessments in clinical settings. We have established a quantitative framework linking epigenetic aging pace to frailty progression and economic outcomes, estimating the costs and benefits of implementing this strategy by comparing potential savings against current projections of aging-related healthcare spending.

Our modeling indicates that intervention scenarios targeting DunedinPACE control result in significant savings compared to the baseline scenario, both in terms of healthcare costs and quality of life improvements. These savings are more pronounced in the long term, as intervention effects accumulate over time to delay frailty onset and progression. The return on investment appears potentially favorable from both healthcare system and individual perspectives, though specific values depend on implementation parameters and economic assumptions.

A key limitation of our current approach is its focus specifically on caloric restriction effects on the frailty pathway. This represents just one facet of a multidimensional relationship between aging interventions and age-related pathologies. Future research should expand this framework to encompass other intervention modalities and specific disease trajectories, particularly neurological, cardiovascular, and musculoskeletal conditions, which collectively constitute major components of age-related healthcare burden. Additionally, the cross-cohort extrapolation between the work by Mak et al.~\cite{Mak2023} and the CALERIE trial introduces uncertainty that should be addressed through further validation studies.

The primary contribution of this work lies in establishing a translational bridge between molecular-level aging biomarkers and economic healthcare metrics via the frailty pathway, demonstrating how emerging technologies in aging assessment might facilitate earlier, more effective intervention strategies. By shifting the intervention window approximately 15 years earlier than conventional aging trials, this approach could fundamentally alter how we conceptualize and address age-related health challenges.

As population aging accelerates globally, implementing evidence-based strategies to extend healthspan becomes increasingly urgent. Our findings suggest that epigenetic pace-of-aging assessments may offer a valuable tool in this endeavor, potentially transforming aging from an inevitable decline into a modifiable process amenable to targeted intervention.

%% include the bibliography 
\bibliography{references_longevity}

\end{document}